\documentclass[aps,twocolumn,showpacs]{revtex4}
\usepackage{graphicx}

\def\be{\begin{equation}}
\def\ee{\end{equation}}
\def\bfi{\begin{figure}}
\def\efi{\end{figure}}
\def\bea{\begin{eqnarray}}
\def\eea{\end{eqnarray}}
\begin{document}
\voffset=+1truecm
\title{Kinetics of bond formation in cross-linked gelatin gels}
\author{T. Abete$^{(a)}$, E. Del Gado$^{(a)}$, D. Hellio
Serughetti$^{(b)}$, L. de Arcangelis$^{(c)}$\footnote{to whom
correspondence should be addressed: dearcangelis@na.infn.it.}, M.
Djabourov$^{(b)}$, A. Coniglio$^{(a)}$} \affiliation{ $^{(a)}$
Department of Physical Sciences and CNISM,
University of Naples ``Federico II'', 80125 Napoli, Italy\\
$^{(b)}$
Laboratoire de Physique Thermique, ESPCI, 75231 Paris, France\\
$^{(c)}$
Department of Information Engineering and CNISM,
Second University of Naples, 81031 Aversa (CE), Italy
}

\begin{abstract}
In chemical cross-linking of gelatin solutions, two different time
scales affect the kinetics of the gel formation in the experiments.
We complement the experimental study with Monte Carlo numerical
simulations of a lattice model. This approach shows that the two
characteristic time scales are related to the formation of single
bonds cross linker-chain and of bridges between chains. In particular
their ratio turns out to control the kinetics of the gel formation.
We discuss the effect of the concentration of chains. Finally our
results suggest that, by varying the probability of forming bridges
as an independent parameter, one can finely tune the kinetics of the
gelation via the ratio of the two characteristic times.
\end{abstract}

\maketitle 
\section{Introduction}
\begin{figure}
\includegraphics[width=9cm]{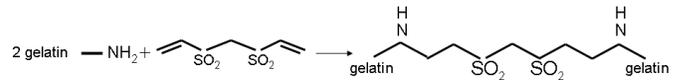}
\caption{Reaction leading to bond formation between an amine group and BVSM}
\label{fig1}
\end{figure}

Among biopolymers, gelatin gels have received great
attention \cite{gelatin} because of their numerous applications in
pharmaceutical, photographic and food industries. When a
semi-diluted gelatin solution is cooled below room temperature, the
coils start to form triple helices and progressively a connected
network is built. The triple helices are reminiscent of the
structure of native collagen, which gave origin to gelatin by means
of a denaturation process. The gel is thermoreversible, i.e. by
raising the temperature the sol state is recovered. Biodiversity due
to chemical composition of the native collagen, molecular weight
distribution, solution properties such as concentration or pH,
influences the temperature of helix formation in the physical gel
\cite{mad1}. The shear modulus shows universal scaling behavior with
a critical exponent $f$ close to 2 versus the distance from the
critical concentration of helices \cite{mad2}. On the other hand,
gelatin solutions show an even richer phenomenology since chemical
gelation or a combination of chemical and physical gelation can be
observed. In fact, if the system is kept above the helix formation
temperature, amino-acids present along the gelatin chain can react
with cross-linking molecules added to the solution. In this case
helices cannot form and a permanent network appears due to
cross-links between reactant and chains, leading to the onset of an
elastic response. Recently, extended studies have been performed on
systems of gelatin in solution with bisvinylsulphonemethyl (BVSM)
reactant \cite{mad3}, able to establish double covalent bonds with
the lysine, the hydroxylysine and possibly with other amine groups
of gelatin chains. The chemical reaction is schematically shown in
Fig.\ref{fig1}.
The influence of various parameters, as gelatin or reagent
concentration and solution pH, on bond formation was investigated.
For instance, increasing the pH activates more amine groups able to
react with BSVM along the gelatin chains. Microcalorimetry
measurements were able to monitor the development of the chemical
reaction in time by detecting the exothermic enthalpy change during
the formation of $C-N$ bonds. Then the kinetics of cross-link
formation was found to follow a double exponential decay with two
characteristic times, whereas a simple exponential decay was
detected at low pH.

However, when counting the number of cross-links binding BVSM and
gelatin, the method could not discriminate between bonds established
by free reactants with a chain and bonds leading to a bridge between
two gelatin chains, nor else bonds leading to a loop within a chain.
This lack of information on the kinetics leading to the gel
formation crucially affects the characterization of the gel
structure and therefore its mechanical properties. Finally it may
reflect onto the location of the gelation threshold and the
determination of the critical exponent of the elastic modulus (the
critical behavior of the shear modulus was measured at low
frequency, giving a critical exponent $f = 3.4 \pm 0.3$\cite{mad3},
close to the expected value for the vulcanization of long chains).
As a consequence, a deeper comprehension of the bond formation
kinetics is essential, requiring alternative investigations. In
particular the primary question is to understand how the two time
scales controlling the kinetics depend on the formation of
single-bonds and bridges between the cross-linkers and the chains,
or else to loops within the chains. Moreover it would be crucial to
understand how these time scales are related to the properties of
the gelatin solution (concentration, pH...) and of the cross-linking
molecules (concentration, reactivity...).

In this paper we analyze the role of concentration and reactivity of
cross-linking molecules in the kinetics of bond formation, by
complementing the experimental observations with a numerical study.
In gelling systems numerical approaches to the study of rheological
and dynamical properties have revealed to be extremely useful for a
better understanding of experimental data \cite{gel_sim,ema,eich}.
Both Monte Carlo and molecular dynamics simulations have been
applied in the last years to the study of different chemical
gelation processes \cite{eich,jullien,nicolai}. Here we have used
Monte Carlo simulations on the cubic lattice of a simple model to
analyze the kinetics of bond formation in chemical cross-linking of
a gelatin solution. We have considered a solution of polymer chains
at different concentrations. Reactant monomers can diffuse in the
solution forming bonds with the active sites along the chains,
producing the cross-linking. Within this approach we have followed
the kinetics of the gel formation varying the gelatin concentration,
the cross-linker concentration and its bonding probability (i.e.
reactivity). Our data reproduce extremely well the experimental
findings. They show that the two time scales detected in the
experiments correspond respectively to the average time of forming
single bonds reactant-chains and bridges chains-chains via
cross-linkers. The ratio of these two characteristic times controls
the kinetics of the bond formation: Varying the concentration and
the cross-linker reactivity strongly affect this ratio and therefore
the kinetics of the gelation process.

The paper is organized as follows: In sect.\ref{exp} the results of
microcalorimetry measurements are presented, whereas in section
\ref{num} the numerical study is described and the results on bond
formation are discussed. Finally in section \ref{bond} the kinetics
of bond formation is analyzed and the concluding remarks are given
in section \ref{conclu}.

\section{Microcalorimetry measurements}
\label{exp} 
\begin{figure}
\includegraphics[width=9cm]{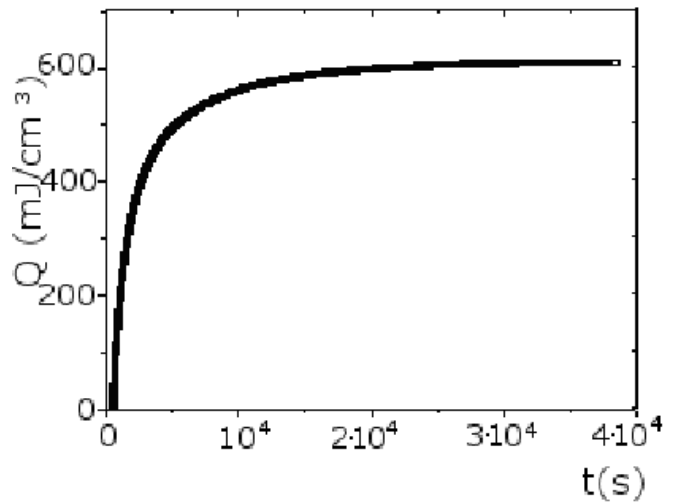}
\caption{Released heat during the chemical reaction between gelatin
and reactant, for a solution with $C_{gel}=12\%$, $C_{BVSM}=0.15\%$,
$pH = 6.7$ and $T=40^oC$}
\label{fig2}
\end{figure}

\begin{figure}
\includegraphics[width=9cm]{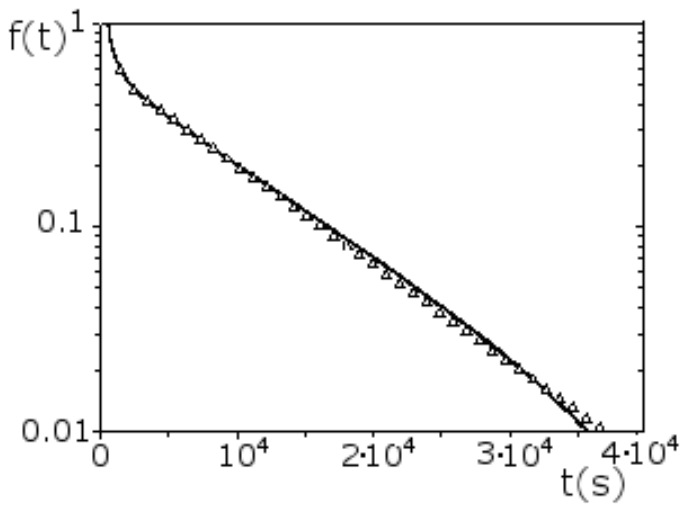}
\caption{
The fraction of bonds of the BVSM that remain to form as a
function of time, for a solution with $C_{gel}=12\%$,
$C_{BVSM}=0.15$, $pH=6.7$ and $T=40^oC$. The continuous line is the
fit with Eq.(\ref{eq:f_exp}).
}
\label{fig3}
\end{figure}

In experiments, the kinetics of the reaction between
amine groups of gelatin chains and BVSM has been monitored by
microcalorimetry measurements. The gelatin sample is a photographic
grade of gelatin extracted from lime processed ossein with an
average molecular weight $M_w\sim 165300 g/mole$, an index of
polydispersity $I_p=2.06$ and an isoelectric point $pI=4.9$. The
granules contain approximately $10\%$ humidity and the
concentrations are corrected accordingly. The BVSM can create
covalent $C-N$ bonds with amine groups of gelatin chains, so that a
permanent network is formed at $T\geq40^oC$, where the triple
helices structure of gelatin gels does not form. D.
Hellio-Serughetti and M. Djabourov \cite{mad3} analyzed the relation
between elastic properties and system parameters was analyzed. The
exothermal reaction between amine groups and BVSM has been monitored
by measuring the enthalpy change. We have performed several
experiments at temperature $T=40^oC$, using solutions with different
concentrations of gelatin and reactant, for different values of pH.
In Fig.\ref{fig2} we plot the released heat $Q(t)$ as a function of
time for a solution with gelatin concentration $C_{gel}=12\%g/cm^3$,
BVSM concentration $C_{BVSM}=0.15\%g/cm^3$ and $pH=6.7$. Normalizing
the released heat by the enthalpy change $\Delta H = - 40 kJ/mol$
due to the formation of one $C-N$ bond, the curve represents at any
time the total number of bonds formed between gelatin chains and
BVSM.
By writing $Q(t) = A (1-f(t))$, where $A$ is a dimensional
coefficient proportional to $\Delta H$, we introduce the function
$f(t)$ which represents the fraction of bonds that remain to form at
time $t$. In Fig.\ref{fig3} $f(t)$ is plotted as a function of time.
Data have been fitted with the sum of two exponentials: \be f(t) =
A_1\exp(-t/\tau_1^m)+A_2\exp(-t/\tau_2^m) \label{eq:f_exp}\ee with
$\tau_1^m=520 s $, $\tau_2^m=9000$$s$ so that
$\tau_2^m/\tau_1^m=17.31$ (the apex "m" is an abbreviation for
"microcalorimetry").
It is worth to notice that microcalorimetry experiments do not allow
to discriminate between single-bonds and bridges or loops within a
chain, where only bridges between different chains contribute to the
increase of connectivity in the system. For this reason computer
simulations are a fundamental step for a deeper understanding of the
kinetics of bonds formation.

\section{Model and numerical study}
\label{num} We have performed Monte Carlo simulations on a cubic
lattice of a system made of bi-functional monomers, i.e. the
reactant and linear chains, which are represented by a sequence of
$n=10$ linked monomers. One monomer of the chain models a Kuhn
segment \cite{Doi}, and therefore represents more units. The length
of a Kuhn segment in a gelatin chain has been measured \cite{Kuhn}
to be of the order of $40$ {\AA}, corresponding to about $10$
amino-acids. As compared to the experiments, our chains correspond
to shorter gelatin chains, containing only about $100$ amino-acids.
Each monomer occupies simultaneously the eight sites of the lattice
elementary cell and, to take into account the excluded volume
interaction, two occupied cells cannot have any site in common. Some
monomers along the chain are active sites which may bind to the
reactant in order to compose complex clusters of chains leading to
the formation of a gel. The active sites are tetra-functional. Two
bonds are formed with the neighbors along the chain and two are not
saturated at the beginning of the simulation. The number of active
sites per chain, $n_{as}$, corresponds to a fixed pH of the
solution. In fact, in experiments the increasing of the pH activates
more amine groups able to react with the BVSM along the chain,
therefore in simulations $n_{as}$ could be varied to study the
effect of the pH. Although the number of amine groups in a gelatin
chain actually linked to reactant cannot be measured experimentally,
it is estimated that at most a fraction of $20\%$ can react.
Therefore we have performed most of the simulations for $n_{as}=5$,
which corresponds to a fraction of $10\%$ of active amino-acids in
our chain.

Chains are randomly distributed on the lattice and diffuse via
random local movements. The excluded volume interaction and the self
avoiding walk condition for polymer clusters restrict the
possibility of monomer movements: to satisfy these  two requirements
the bond lengths vary into a set of permitted values according to
bond-fluctuation dynamics \cite{BFD}. On a cubic lattice the allowed
bond lengths are $l=2,\sqrt5,\sqrt6,3,\sqrt10$. At each Monte Carlo
step the time is increased by $\delta t=1$ and one random move is
selected on average for each monomer: if the move satisfies the
bond-fluctuation dynamics and excluded volume conditions, it is
executed, otherwise it is rejected. These simple laws for local
movements give rise to a dynamics which takes into account the main
features of the real dynamics of polymer molecules \cite{BFD}.

After chains have diffused and reached equilibrium, we add the
reactant to the system and let the solution diffuse towards the
stationary state. Due to the diffusion of cross-linkers and chains,
when a reactant finds a nearest neighbor unsaturated active site, a
bond may form. The process goes on until all the possible bonds are
formed.

The bond formation may request to overcome a free energy barrier
\cite{malla}, depending on the nature of the solution, the active
sites and the reactant. In particular, it may depend on some
specific local orientation of the molecules, some restriction on the
value of the angle between two bonds, due to the rigidity of the
$C-N$ link, or else may be affected by variations of the effective
reactivity of the cross-linker. In our model we have taken into
account these effects in the following way: The first bond of a
reactant monomer is formed along lattice directions as soon as there
is a neighboring active site. The second bond is formed with
probability $p_b\leq1$, since a reactant monomer is expected to have
less chance to react when is already bonded to a chain, compared to
when it is free. In the same spirit of reaction limited aggregation
models \cite{rlca}, $p_b$ is an independent parameter which
influences the time of formation of bridges between gelatin active
sites. The value of $p_b$ should be determined by the features of
cross-linking reagent. Moreover, although varying the bridge
probability $p_b$ does not affect the gelation transition, it has a
crucial effect on the velocity of the reaction, which can be easily
enlightened in the numerical simulation as discussed later.

\begin{figure}
\includegraphics[width=9cm]{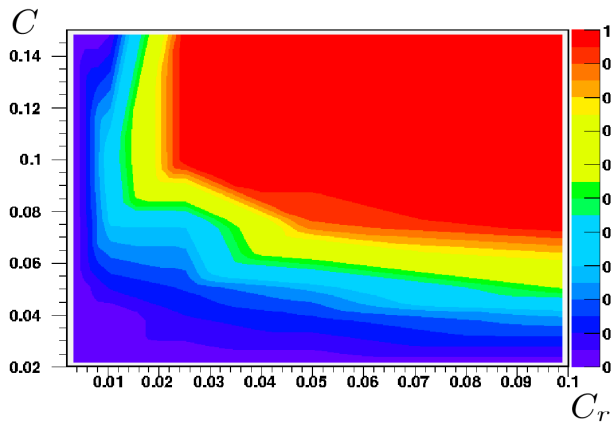}
\caption{
(Color online)
The phase diagram, obtained by plotting the percolation probability $\Pi$
as a function of chain and reactant concentration $C$ and $C_r$,
respectively, using a color scale, reported in legend.
The spanning probability has been averaged over $30$ independent
configurations of a sample of size $L=100$. The
number of active sites per chain $n_{as}=5$ is kept constant and
$p_b=1$. The percolation line can be identified with
the locus $\Pi=0.5$.
}
\label{fig4}
\end{figure}

We have performed numerical simulations of the model for different
lattice sizes ($L=50,100,200$), where the unit length is the lattice
spacing $a=1$, with periodic boundary conditions. The chain
concentration $C$ and the cross-linker concentration $C_r$ are
defined as the ratio between the number of monomers/reactant and the
maximum number of monomers $N_{max}=L^3/8$ in the system. Using the
percolation approach we identify the gel phase as the state in which
there is a percolating cluster, which spans the whole system
\cite{Flory_perco}. For a fixed set of parameters we generate a
number of configurations of the system and monitor the reaction. In
order to locate the gelation transition we analyze the percolation
probability $\Pi$, defined as the fraction of configurations leading
to a percolating cluster, and we identify the transition with the
line $\Pi=0.5$ \cite{stauffer}. We have determined a qualitative
phase diagram (Fig.\ref{fig4}) by varying the chain and cross-linker
concentrations, $C$ and $C_r$ respectively, for a fixed $n_{as}=5$.
In experiments the total amount of reactant has been consumed at the
end of the reaction process, i.e. the amount of reactant is much
lower than the amount of active sites. The reaction stops when all
the reactant are linked to amine groups and the system is in the gel
phase. Moreover the experimental system \cite{mad3} is investigated
at gelatin concentrations $C_{gel}$ above the overlap concentration
$C_{gel}^*=0.005$ $g/cm^{-3}$. In order to reproduce the
experimental conditions the crosslinker concentration has been fixed
at $C=0.025$, which corresponds to $C_r\ll C$ at the sol-gel
transition. Under this condition, in simulations the gel phase is
located at concentrations $C$ above the overlap concentration which
in our system is $C^*\simeq 0.017$. This choice of parameters
guarantees that at the end of the reaction the system is in the gel
phase.

With finite size scaling analysis we have obtained the percolation
threshold, that for the case $C_r=0.025$ is $C_c=0.10\pm0.05$, the
critical exponents $\nu=0.9\pm0.1$ for the percolation connectivity
length $\xi$ ($\sim |C-C_c|^{-\nu}$) and $\gamma=1.78\pm0.10$ for
the mean cluster size $\chi$ ($\sim |C-C_c|^{-\gamma}$). These
results are in good agreement with the random percolation critical
exponents\cite{stauffer}. The random percolating cluster is
characterized by a fractal structure: its mass $M$, i.e. the number
of monomers, scales with its linear dimension $\xi$ with a power law
behavior $\xi^D$, where $D\simeq 2.5$ \cite{stauffer}. The structure
of the formed network may depend on model parameters and influences
the rheological response of the system.

\section{Kinetics of bond formation}
\label{bond} In simulations we have investigated the behavior of the
number of bonds formed during the reaction process and we have
distinguished between: \\1. Bonds between a free reactant and an
active site (we will refer to the latter type of bond as
\emph{single-bonds}); \\2. Bonds between a linked reactant and an active site
of another chain;
\\3. Bonds between a reactant and two active sites of the same chain
(which in the following we will call \emph{loops}).
\\We have analyzed the kinetics of bond formation for a system of size
$L=100$, with $C_r=0.025$ and $n_{as}=5$ varying the chain
concentration $C$ and the probability $p_b$ of bridge formation. The
time is measured in Monte Carlo unit time $\delta t$.

\begin{figure}
\includegraphics[width=9cm]{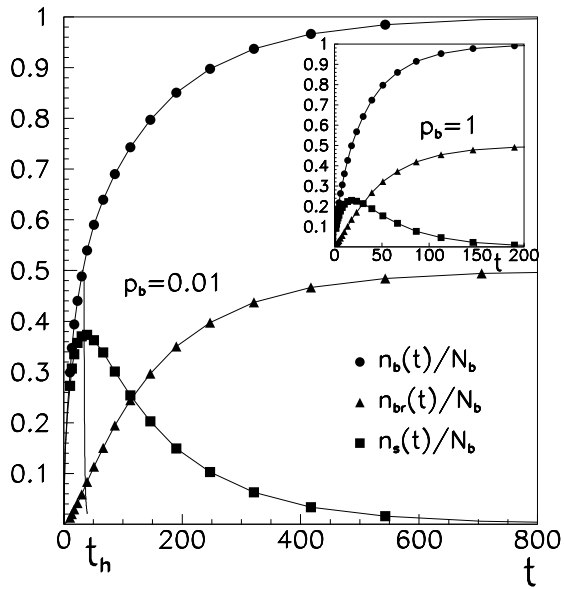}
\caption{
Total number of bonds, number of bridges and number of
single-bonds normalized by the total number of possible bonds $N_b$
as a function of time for $C=0.3$, $C_r=0.025$, $n_{sa}=5$ and
$p_b=0.01$. Inset: The same quantities for $p_b=1$. $t_h$ indicates
the time corresponding to half of the reaction.
}
\label{fig5}
\end{figure}

\begin{figure}
\includegraphics[width=9cm]{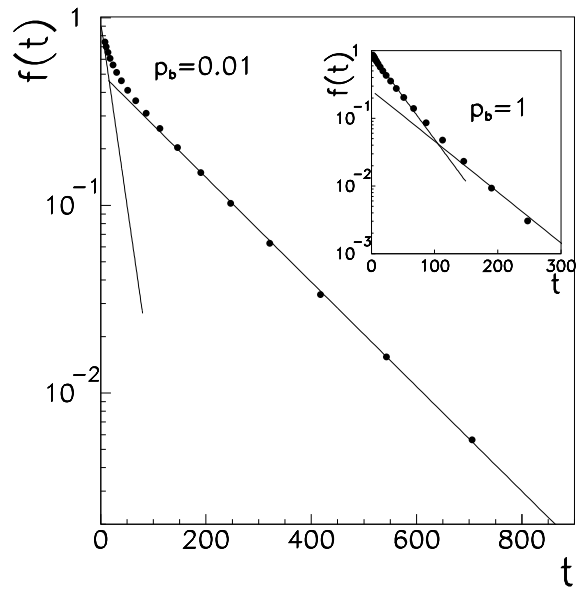}
\caption{
Function $f(t)=1-n_{b}(t)/N_b$ as a function of time, where
$N_b$ is the total number of possible bonds, for $C_r=0.025$,
$n_{as}=5$ and $p_b=0.01$. The full lines are the fitting curves
$\sim exp(-t/\tau_1)$ with $\tau_1=20$, and $\sim exp(-t/\tau_2)$
with $\tau_2=166$. Inset: $f(t)$ with $p_b=1$. The full lines are
the fitting curves $\sim exp(-t/\tau_1)$ with $\tau_1=34.4$ and
$\sim exp(-t/\tau_2)$ with $\tau_2=57.8$.
}
\label{fig6}
\end{figure}

Since $C_r \ll C$, the total number of bonds at the end of reaction
is equal to twice the number of cross-linkers: $N_b=2 C_r L^3/8$. In
Fig.\ref{fig5} the total number of bonds $n_b(t)$ is plotted as a
function of the time together with the number of single-bonds
$n_{s}(t)$ (bonds of type 1) and bridges $n_{br}(t)$ (i.e. bonds of
type 2 or 3) in the case $p_b =0.01$. The number of bonds has been
normalized by $N_b$. As the reaction begins, single-bonds form
rapidly, then bridges start to form and the degree of connectivity
between chains increases. The behavior of the total number of bonds
versus time closely resembles the released heat experimentally
measured during the reaction reported in Fig.\ref{fig2}. The
velocity of the reaction is related to the probability $p_b$ of
bridge formation. It governs the mean time of link formation between
different chains and strongly influences the duration of the
reaction process. In the inset of Fig.\ref{fig5} $n_b(t)$, $n_s(t)$
and $n_{br}(t)$ are plotted as a function of the time for $p_b=1$,
showing that both single-bonds and bridges form more rapidly as
compared to the $p_b=0.01$ case.
The total number of bonds $n_b(t)=n_s(t)+2n_{br}(t)$, and its time
dependence can be written as $n_b(t)=\!N_b(1-f(t))$. In
Fig.\ref{fig6}, the function $f(t)=1-n_{b}(t)/N_b$, representing the
fraction of bonds that remain to form, is plotted in a semi
logarithmic plot for the case $p_{b}=0.01$. The data reproduce
extremely well the behavior observed experimentally (Fig.\ref{fig3})
and are well fitted by a sum of two exponentials
\begin{equation}
f(t) = a_1\!\cdot\! e^{(-t/\tau_1)} + a_2\!\cdot\!
e^{(-t/\tau_2)}\!
\label{nb}
\end{equation}
in agreement with the microcalorimetry measurements
(Eq.\ref{eq:f_exp}). From the fit we obtain $\tau_1=20\pm2$ and
$\tau_2=166\pm5$, for $p_b=0.01$ and $C=0.3$, providing
$\tau_2/\tau_1=8.3\pm0.9$.
If the bridge probability $p_b$ varies, the mean time of bridge
formation $\tau_2$ changes, and so does the ratio $\tau_2/\tau_1$.
In the inset of Fig.\ref{fig6} $f(t)$ is plotted for $p_b=1$. From
the fit we obtain $\tau_1=34.4\pm1.5$, $\tau_2=57.8\pm1.5$, and
hence $\tau_2/\tau_1=1.68\pm 0.08$.

In order to give a microscopic interpretation for these two
characteristic times, we have directly computed the average time of
formation of single bonds and bridges respectively, which cannot be
done by microcalorimetry measurements. These two times are in
agreement with the fitting parameters $\tau_1$ and $\tau_2$ within
error bars. Next, we have analyzed the mean square displacement of
the reactant monomers when they are free (1) and when they have
formed one single bond (2). Interestingly we have observed that the
ratio between the corresponding diffusion coefficients $D_1/D_2$ is
of the order of the ration $\tau_2/\tau_1$ for all the analyzed
concentrations of chains and cross-linkers. This result suggests
that $\tau_1$ and $\tau_2$ are related to the characteristic times
for diffusion of the free reactant and of the reactant linked to a
gelatin chain, respectively.

In agreement with this microscopic interpretation, our data
(Fig.\ref{fig5} and \ref{fig6}) show that, for the concentrations
$C$ and $C_r$ explored, the single-bonds form more rapidly than
bridges. This different velocity of formation is due, apart from
$p_b$, to the different mobility of free cross-linkers with respect
to linked ones, which are forced to move together to the chains to
which they are permanently bonded. As a consequence,
$\tau_1\leq\tau_2$, i.e. the average time of formation of a
single-bond is generally smaller than the average time of formation
of bridges even for $p_b=1$.

\begin{figure}
\includegraphics[width=9cm]{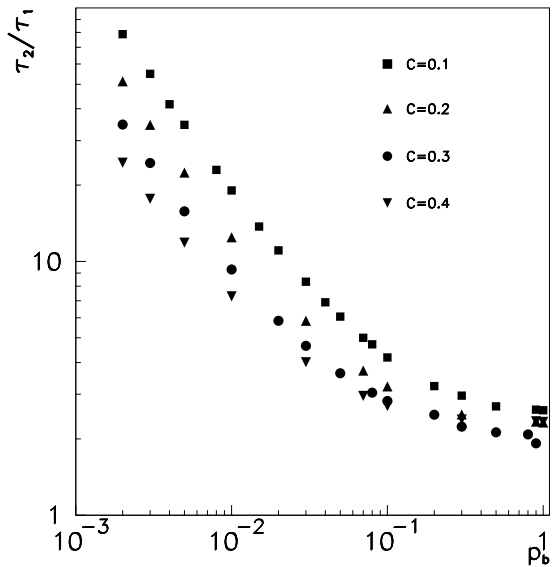}
\caption{
The ratio $\tau_2/\tau_1$ between the average time of
formation of bridges and the average time of formation of
single-bonds as a function of bridge probability $p_b$ for different
concentrations $C$ of chains, for $C_r=0.025$ and $n_{as}=5$.
}
\label{fig7}
\end{figure}

It is interesting to notice that in simulations we can easily vary
the bridge probability $p_b$, to see how the features of the
cross-linking reagent could possibly affect the kinetics of the bond
formation. Remarkably, we find that this effect can be crucial.
Indeed, as $p_b$ governs the bridges formation, it influences the
average time $\tau_2$: As $p_b$ decreases, the reaction slows down
and $\tau_2$ increases. We have systematically analyzed the behavior
of $\tau_2/\tau_1$, as a function of bridge probability $p_b$. In
Fig.\ref{fig7} we plot the obtained data: When the bridge
probability $p_b$ increases, the average time of bridge formation
decreases, and so does the ratio $\tau_2/\tau_1$. For $p_b\gtrsim
0.3$ we find that the ratio decreases more slowly apparently tending
to a plateau. In this regime, the bridge probability $p_b$ does not
influence the kinetics of bond formation, which is completely
governed by the diffusion of the monomers and by their
concentration. The value of the ratio between the two characteristic
times obtained in the experiments at $pH =6.7$ (see Fig.\ref{fig3})
corresponds to a reactant with $0.005\lesssim p_b \lesssim0.01$ in
our simulations.
The experimental findings show that the solution pH also
affects the kinetics of bond formation, i.e. the ratio
$\tau_2^m/\tau_1^m$ decreases as the pH decreases. 
This is in agreement with the chemistry of reaction, where the non
protoned form of the amine is reactive. In fact, one could expect
different regimes depending on the chains concentration. Indeed, if
decreasing of the number of active sites will in general correspond
to an increase of $\tau_1$, the effect of this variation on $\tau_2$
is likely to strongly depend on chain concentration. At high
concentration of chains, if the number of active sites per chain
decreases below a certain level, we expect that these will be
surrounded by many other sites of the same chain which are not
active. As a consequence, due to excluded volume effects, they will
hardly be reached by partially linked cross-linkers, i.e. $\tau_2$
will increase as the number of active sites decreases and so will
the ratio $\tau_2/\tau_1$. On the other hand, $\tau_2$ strongly
depends on the chain mobility and on the formation of loops.
Therefore decreasing the pH at low concentrations, leads to an
increase of $\tau_1$ that may be balanced by a not so dramatic
increase of $\tau_2$, due to the eventual formation of loops. As a
consequence one can observe a net decrease of the ratio
$\tau_2/\tau_1$. We would like to stress that, although the
investigation of the role of pH could be in principle done with this
model by varying $n_{as}$, one should use long enough chains to be
able to detect the different concentration regimes.
\\Conversely, when the number active sites increases up to $n_{as}=10$,
in the analyzed concentration range the average time $\tau_1$ does
not change appreciably, while $\tau_2$ decreases due to the
formation of loops. For bridge probability $p_b$ sufficiently high
($p_b \gtrsim 0.8$), the mean time of bridge formation becomes less
than or equal to the mean time of formation of single-bonds: in
these cases, the number of bonds $n_b(t)$ may be fitted by a single
exponential with the characteristic time $\tau_1$.

\begin{figure}
\includegraphics[width=9cm]{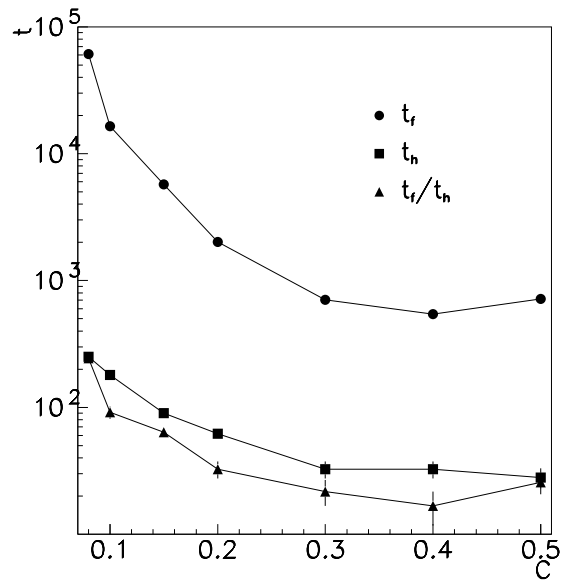}
\caption{
The time of reaction end $t_f$ (dots), time of half
reaction $t_h$ (squares) and their ratio (triangles) as a function
of chain concentration $C$ for $C_r=0.0025$, $n_{as}=5$ and
$p_b=0.01$.
}
\label{fig8}
\end{figure}

\begin{figure}
\includegraphics[width=9cm]{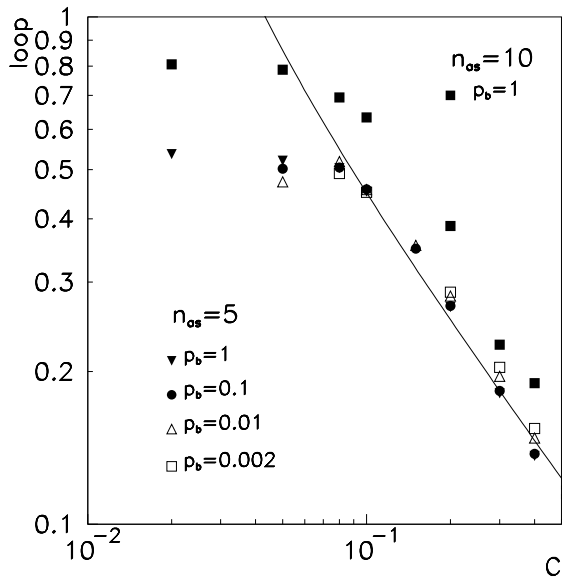}
\caption{
Number of loops normalized by the maximum number of
bridges $N_b/2$, as a function of gelatin concentration $C$ for
$p_b=0.002,0.01,0.1,1$ for $n_{as}=5$ and for $p_b=1$ with
$n_{as}=10$, for $C_r=0.025$. The full line is the power law fitting
curve $\sim C^{-l}$ with $l=0.75$.
}
\label{fig9}
\end{figure}

To complete our study of the kinetics of bond formation, we have
measured the time $t_h$ of formation of half of the total bonds, and
the duration of the reaction $t_f$, i.e. the average time needed to
form all possible bonds. The data are presented in Fig.\ref{fig8}
and show that the ratio $t_f/t_h$ decreases as the concentration
increases, tending to a plateau value for $C \gtrsim 0.3$. This
behavior is in agreement with the experimental findings: For
$C_{BVSM}=0.3\%$ $g/cm^{3}$ and $C_{gel}$ ranging from $3 \%$ to $6
\%$, the ratio $t_f/t_h$ decreases from $t_f/t_h \sim 27$ to
$t_f/t_h\sim 15$. For $C_{gel} \gtrsim 6 \%$ the ratio remains
constant with the concentration. Comparing the experimental results
with simulations, we conclude that the regime of concentration
$C\sim 0.3$ for the lattice model corresponds to $C_{gel}\sim 6\%$
of the experiments \cite{mad3}. This correspondence is coherent with
the behavior shown in Fig.s \ref{fig5},\ref{fig6},\ref{fig7} and
support the suggested interpretation. It is worth to notice that
$t_f/t_h\gg1$ for the explored range of parameters: this is a
consequence of the fact that in the first half of the reaction most
of formed bonds are single-bonds. In the second half, bridges
between active sites are established, requiring a longer time.
Finally we have analyzed the number of loops which are formed during
the cross-linking reaction and may play a crucial role in the
mechanical response of the gel. Loops are not detectable in
experimental measurements, but can be easily monitored in
simulations. Previous numerical simulations of polymerization
process in hexamethylene diisocynate-based polyurethane
\cite{Stepto} indicate that the number of loops has different roles
in the various concentration regimes. Indeed the loss of elasticity
due to loops may be outweighed by the increase of topological
entanglements, depending on the concentration.
\\At the end of the reaction we have counted the number of loops,
normalized by the maximum number of bridges $N_b/2$, and
investigated its behavior as a function of $C$. Data plotted in
Fig.\ref{fig9} refer to chain with $n_{as}=5$ and $n_{as}=10$. Our
results indicate that the number of loops decreases as the chain
density $C$ increases. In the range of concentration explored the
number of loops decreases following a power law behavior $\sim
C^{-l}$ characterized by an exponent $l=0.75\pm0.05$. The behavior
appears independent of the bridge probability $p_b$: This result
confirms that the bridge probability only influences the kinetics of
bond formation but does not strongly affect the morphology of the
system. By opportunely tuning $p_b$, i.e. changing the reactant, the
velocity of the reaction may be adjusted and the formation of
single-bonds and bridges may be tuned in time, but the final
geometrical properties of the structure should not be modified. On
the other hand, the connectedness of the system may be influenced by
the number of active sites per chain $n_{as}$. In particular, when
$n_{as}$ increases, the number of loops formed in the system
increases and hence the degree of connectedness of the system
decreases (Fig.\ref{fig9}). Moreover, in the limit of very diluted
solutions of chains where all sites are active ($n_{as}=10$), loops
represent about $80\%$ of the total number of bonds and, as a
consequence, the viscoelastic properties can be sensibly modified
\cite{Stepto}.

\section{Conclusion}
\label{conclu} In conclusion, our experimental results show that two
different timescales affect the kinetics of bond formation in our
cross-linked gelatin solution. The numerical data reproduce well the
experimental ones and clarify the mechanisms involved in bond
formation. Our study shows that the two time scales detected in
experiments correspond to the average time of forming single bonds
reactant-chains and bridges chains-chains via cross-linkers. These
two times are related to the characteristic times of diffusion of
free reactants and reactants which have already formed one bond.
Their ratio controls the kinetics of the bond formation: Varying the
concentration, the cross-linker reactivity and the pH strongly
affect this ratio and therefore the kinetics of the gelation
process. Our findings also show that the probability $p_b$ to form a
bridge between two active sites allows to finely tune the kinetics
of the reaction via the ratio of the two characteristic times. A
variation of $p_b$ in our interpretation corresponds to a variation
of the free energy barrier to be overcome in order to form the bond,
or to different orientations of bonds vectors; hence to vary $p_b$
corresponds to change the reactant agent in the gelatin
solution. 
Moreover, our data indicate that the number of loops formed between
two active sites of the same chain, which has an important effect of
the viscoelastic properties of the system, increases when the pH of
the solution increases. This model represents a useful tool to
investigate rheological behavior of gelatin solutions and the
relation between the kinetics and gel structures.

Acknowledgements. This work was supported by MIUR-PRIN 2004,
MIUR-FIRB 2001, CRdC-AMRA, the Marie Curie Reintegration Grant
MERG-CT-2004-012867 and EU Network Number MRTN-CT-2003-504712.


\end{document}